\documentclass{elsart}
\usepackage{epsfig}
\usepackage{amssymb}
\def\bea {\begin{eqnarray}}
\def\eea {\end{eqnarray}}

\def\be {\begin{equation}}
\def\ee {\end{equation}}
\begin{document} 

\begin{frontmatter} 

\title{Characterizing quark gluon plasma by dilepton interferometry}

\author{Payal Mohanty, Jan-e Alam and Bedangadas Mohanty}
\medskip
\address{Variable Energy Cyclotron Centre, 1/AF, Bidhan Nagar, 
Kolkata - 700064, INDIA}
\begin{abstract}
The Hanbury-Brown-Twiss (HBT) radii have been calculated from
the two particle correlation functions with virtual photons 
produced in the collisions
of two nuclei at ultra-relativistic energies. We show that the variation of
the HBT radii with the invariant mass of the virtual photon can be
used to characterize and distinguish the hadronic as well as the partonic 
phase that might have produced initially in the collisions.  
It has been  illustrated that the non-monotonic 
variation of the HBT radii with invariant mass
provides an access to the development of collective flow  in the system.
\end{abstract}


\end{frontmatter} 

The aim of the nuclear collisions at ultra-relativistic energies 
is to create and study - a state of  matter,
where quarks and gluons are dislocated from individual hadrons and 
make an excursion over a nuclear volume - such a phase of matter
is called quark gluon plasma (QGP). Several probes - both electromagnetic (EM)
and hadronic  have been proposed for the diagnostics of QGP. 
The electromagnetically interacting probes (real and virtual 
photons) has the advantage over the
hadronic probes  because of their nature of interaction. The EM probes
has mean free path much larger than the size of the system as a consequence
they leave the system without re-scattering and hence 
can transmit the source information  very efficiently~\cite{lm}. 
However, photons and lepton pairs can be produced 
from both the partonic as well as hadronic phases~\cite{ja}.
As a consequence the disentanglement of the contribution 
for the QGP still remains a big challenge.

In case of EM probes- dilepton has the advantage over the real photons.
The photons with low transverse momentum ($k_T$) from the 
hadronic phase may receive large transverse kick
due to  radial flow and consequently appear as high $k_T$ photons. 
These photons can mingle with  the 
the contributions from the high temperature QGP phase,
making the detection of photons from QGP difficult. 
However, for dileptons there are two kinematic variables available - the
$k_T$ and the invariant mass ($M$).  
While the $k_T$ spectra of dilepton is affected by the flow, 
the $k_T$ integrated $M$ spectra remains unchanged.
This suggests that a careful selection of $k_T$ and
$M$ windows will be very useful to characterize the
QGP and the hadronic phases.

The interferometry of the dilepton pairs actually reflect correlation
between two virtual photons, the analysis then concentrates on
computing the Bose-Einstein correlation (BEC) function for two 
virtual photons which can be defined as,
\begin{equation}
C_{2}(\vec{k_{1}}, \vec{k_{2}}) = 
1+\frac{[\int d^4x~\omega(x,K)\cos(\Delta\alpha)]^2+
[\int d^4x~\omega(x,K)\sin(\Delta\alpha)]^2}{P_1(\vec{k_1})P_1(\vec{k_2)}}
\label{eq1}
\end{equation}
where $k_i$ is momentum of the individual photon, 
$K=(k_1+k_2)/2$, 
$\Delta\alpha=\alpha_1-\alpha_2$,  
$\alpha_i=\tau M_{iT}\cosh(y_i-\eta)-r k_{iT}\cos(\theta-\psi_i)$,
$M_{iT}=\sqrt{k_{iT}^2+M^2}$ is the transverse mass and
$\omega(x,K)$ is the source function related to the thermal emission rate 
of the virtual photons  per unit four volume 
(see ~\cite{hbt,hbtlep} for details).

For the space time evolution of the system relativistic hydrodynamical
model with cylindrical symmetry~\cite{hvg} and  boost invariance along
the longitudinal direction~\cite{jdb} has been used. 
For a system undergoing isentropic expansion,
the initial 
temperature ($T_{i}$) and proper thermalization time 
($\tau_{i}$) of the system 
may be constrained by the measured hadronic multiplicity,
$dN/dy\sim T_i^3\tau_i$. 
Here we have taken $T_i=290$ MeV and $\tau_i=0.6$ fm/c.
The equation of state (EoS) which controls the rate of expansion/cooling
has been taken from the lattice QCD calculations ~\cite{MILC}. 
The chemical ($T_{ch}$=170 MeV) and kinetic ($T_{fo}$=120 MeV)
freeze-out temperatures 
 are fixed by the particle ratios and the slope 
of the $k_T$ spectra of hadrons~\cite{hirano}. 
With all these ingredients the correlation function $C_2$ 
has been  evaluated for different invariant mass windows 
($\langle M \rangle=(M_1+M_2)/2$=0.3, 0.5, 0.7, 1.2, 1.6 and 2.5 GeV) 
as a function of $q_{side}$ and $q_{out}$~\cite{hbt} which are related 
to transverse momentum of individual pair.
Once the correlation function 
for the (time like) virtual photon is calculated, the source dimensions can be 
obtained by parameterizing it with the empirical (Gaussian) form:
\be
C_2=1+\frac{\lambda}{3}\exp(-R^2_{i}q^2_{i}).
\label{eq2}
\ee
where the subscript $i$ stand for $out$ and $side$.  In Eq.~\ref{eq2} $\lambda$ 
represents the degree of chaoticity of the source.  Deviation of $\lambda$ from
1 will indicate the presence of non-chaotic sources.

\begin{figure}[h]
\begin{center}
\includegraphics[scale=0.4]{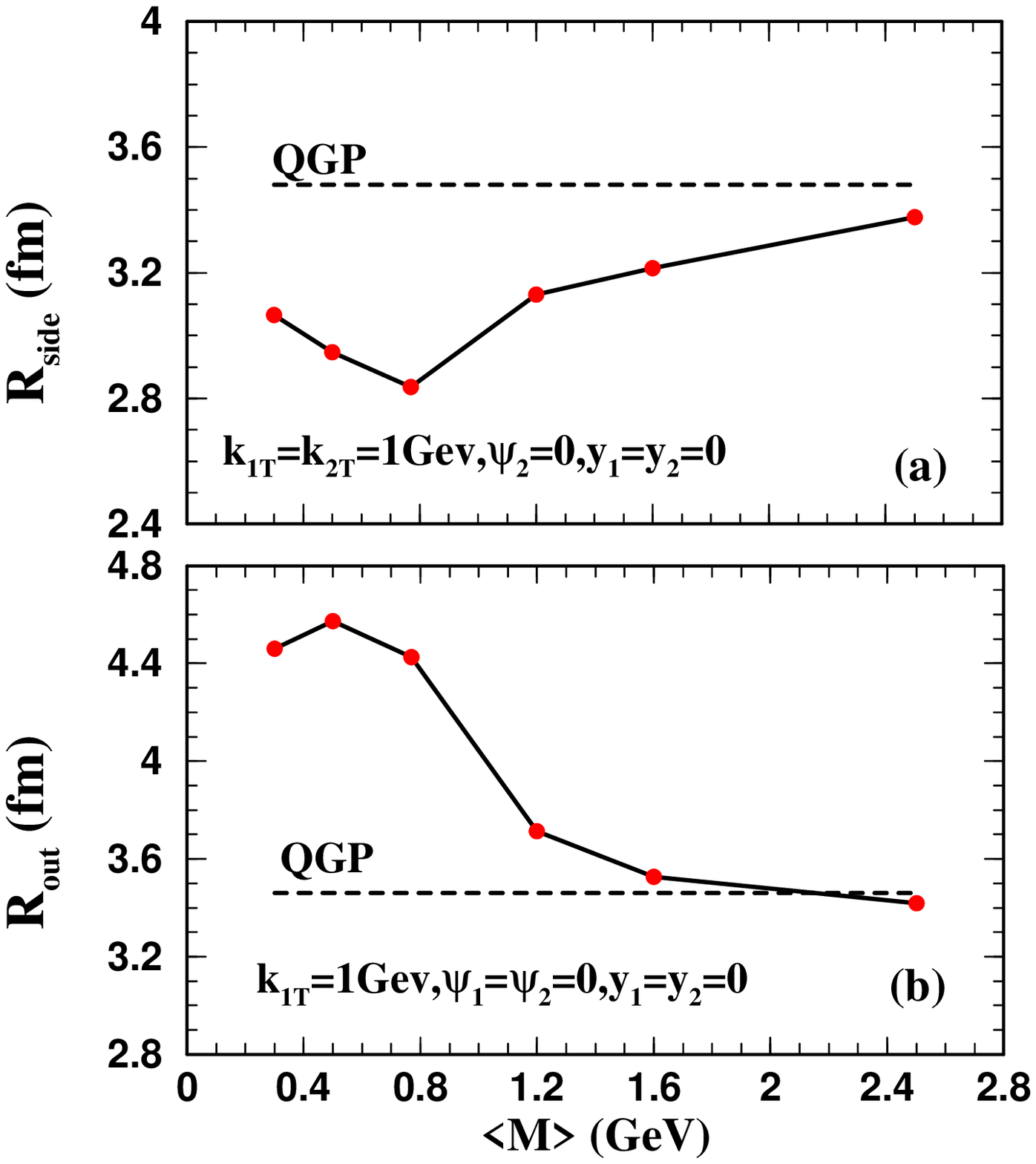}
\includegraphics[scale=0.4]{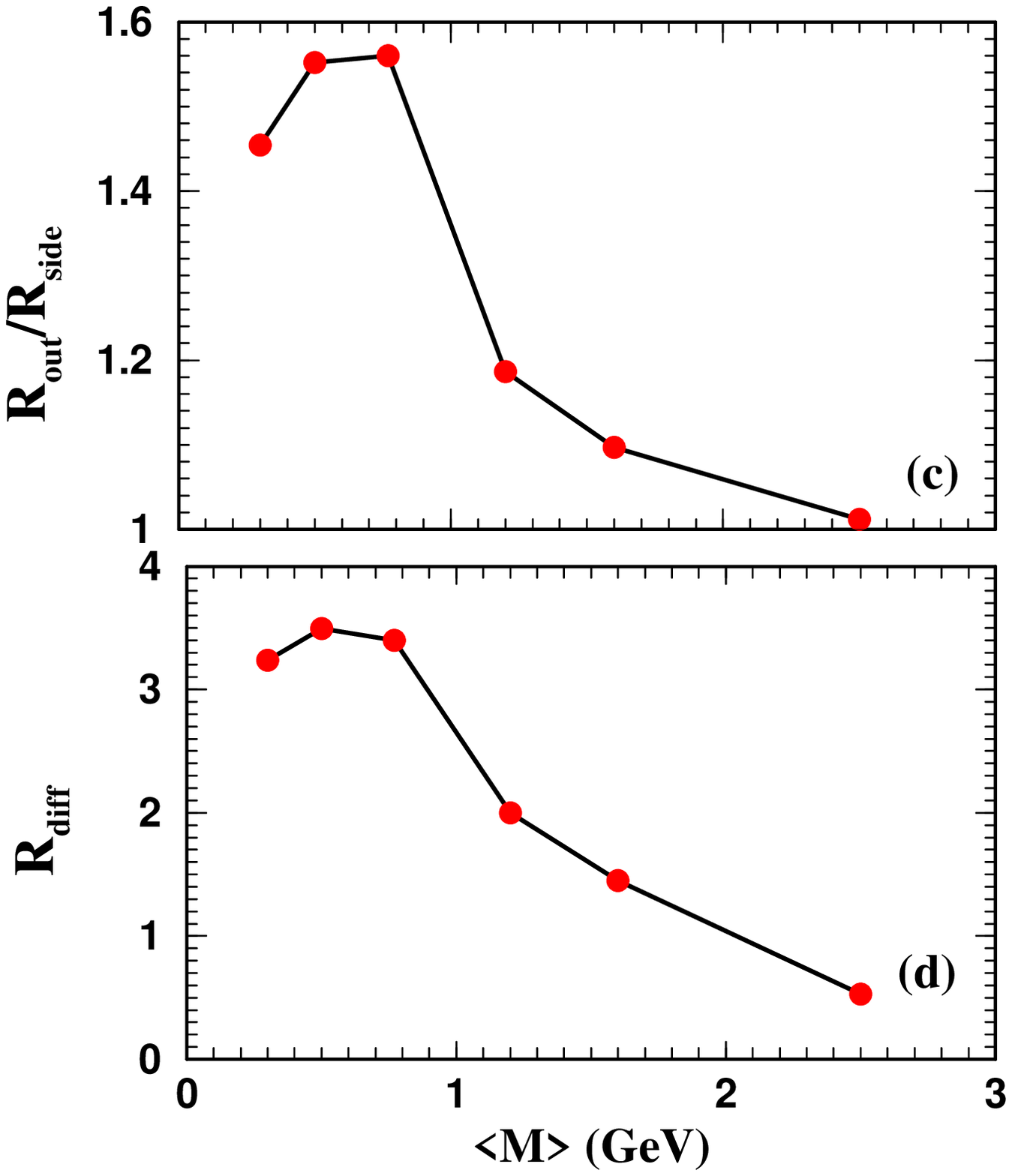}
\caption{The variation of 
$R_{side}$ and $R_{out}$ as a function of $\langle M \rangle$ (left panel).  
The right panel show $R_{out}/R_{side}$ and
$R_{diff}=\sqrt{R_{out}^2-R_{side}^2}$ as a function of $\langle M\rangle$.
}
\label{fig1}
\end{center}
\end{figure}

The $R_{\mathrm side}$, radius corresponding to $q_{side}$ 
is closely related to the transverse size of the system and 
considerably affected by the collectivity and the
$R_{\mathrm out}$, radius corresponding to $q_{out}$ measures both the 
transverse size and duration of particle emission ~\cite{hb3,uaw}. 
The $R_{\mathrm side}$ shows non-monotonic dependence on $\langle M\rangle$
(Fig.~\ref{fig1}a). It can be shown
that $R_{side}\sim 1/(1+E_{\mathrm collective}/E_{\mathrm thermal})$. 
With the transverse expansion the radial size of the emission zone 
decreases with time as  
a rarefaction wave moves toward the center of the cylindrical geometry.
The high $\langle M\rangle$ regions 
are dominated by the early partonic phase~\cite{hbt} 
where the collective flow has not been developed fully~\cite{pmprc} 
consequently the ratio of collective to thermal energy is small-
hence a larger $R_{\mathrm side}$ is obtained for large $M$.
In contrast, the lepton pairs with $M\sim m_\rho$ 
are emitted from the late hadronic phase where the size of the emission zone 
is smaller due to larger collective flow giving rise to
a smaller $R_{\mathrm side}$. The ratio of collective to thermal 
energy for such cases is quite large, which is reflected as a dip
in the variation of $R_{\mathrm side}$ with $\langle M\rangle$ 
around the $\rho$-mass region (Fig.~\ref{fig1}a). 
Thus the variation of $R_{\mathrm side}$
with $M$ can be used as an efficient tool to measure the
collectivity in various phases of matter. 
The dip, in fact vanishes if the contributions from $\rho$ and $\omega$ 
is switched off~\cite{hbtlep}. 
We observe that by keeping the $\rho$ and $\omega$ contributions
and setting radial velocity, $v_r=0$, the dip in $R_{\mathrm side}$
vanishes, confirming
the fact that the dip is caused by the 
large radial flow of the hadronic matter~\cite{hbtlep}.   
The $R_{\mathrm out}$ probes both the transverse dimension as well as the 
duration of emission and unlike $R_{\mathrm side}$, $R_{\mathrm out}$ does not
remain constant even in the absence of radial flow. The large $M$ regions are
populated by lepton pairs from early partonic phase where the
effect of flow is small and the duration of emission is also
small - resulting in smaller values of $R_{\mathrm out}$. 
For lepton pair from $M\sim m_\rho$ region  the flow is large
which could have resulted in a dip as in $R_{\mathrm side}$ in
this $M$ region. However, $R_{\mathrm out}$ probes the duration
of emission too which is large for hadronic phase.
The larger duration overwhelms
the reduction of $R_{\mathrm out}$
due to flow in the hadronic phase
resulting in a bump in $R_{\mathrm out}$ in this region of $M$
(Fig.~\ref{fig1}b). 

Figs.~\ref{fig1}(c) and 1(d) show the variation of $R_{out}$/$R_{side}$ and  
$R_{diff}=\sqrt{R_{out}^{2}-R_{side}^{2}}$  with
$\langle M \rangle$.
These two quantities provides information on the duration of particle 
emission~\cite{uaw} 
for various domains of $M$. In one hand the high and low $M$ domains  
are dominated by radiation from early QGP phase~\cite{hbtlep}
where the duration of 
particle emission is small. 
On the other hand the lepton pairs for $M$ around $\rho$ mass dominantly
originate from the late hadronic phase where the duration of particle 
emission is large, Resulting in a
non-monotonic variation of $R_{diff}$, 
which indicates the presence two different phases during the evolution.

In summary, we have evaluated the dilepton pair correlation functions 
relevant for Au+Au collisions at RHIC energy. 
The values of HBT radii extracted from the dilepton correlation 
functions show non-monotonic dependence on dilepton pair mass,
reflecting the evolution of collective flow in the system. 
It appears that the non-monotonic variation of 
$R_{\mathrm out}$ and $R_{\mathrm side}$ with $M$ 
originate from the presence of two phases during the evolution of
the system, hence could be use as a signal of phase transition.

{\bf Acknowledgment:} We are grateful to Tetsufumi Hirano for providing
us the hadronic chemical potentials. 
We thank Nu Xu for very useful discussions. 
J A and P M supported by DAE-BRNS project Sanction No. 2005/21/5-BRNS/2455  
and B M supported by DAE-BRNS project Sanction No. 2010/21/15-BRNS/2026.

\end{document}